# Real-time Assessment of Right and Left Ventricular Volumes and Function in Children Using High Spatiotemporal Resolution Spiral bSSFP with Compressed Sensing

Jennifer A. Steeden (PhD, jennifer.steeden@ucl.ac.uk)[1],

Grzegorz T. Kowalik (PhD, grzegorz.kowalik.09@ucl.ac.uk)[1],

Oliver Tann (oliver.tann@gosh.nhs.uk)[2],

Marina Hughes (marina.hughes@gosh.nhs.uk)[2],

Kristian H. Mortensen (PhD, Kristian.Mortensen@gosh.nhs.uk)[2],

Vivek Muthurangu (MD, v.muthurangu@ucl.ac.uk)[1*]

1. *UCL Centre for Cardiovascular Imaging, University College London, London. WC1N 1EH. United Kingdom*
2. *Cardiorespiratory Unit, Great Ormond Street Hospital for Children, London. WC1N 3JH. United Kingdom*

**Corresponding Author:** Vivek Muthurangu

UCL Centre for Cardiovascular Imaging, Institute of Cardiovascular Science

30 Guildford Street, London. WC1N 1EH

Tel: +44 (0)207 762 6834

Fax: +44 (0)207 813 8263

# ABSTRACT

**Background**

Real-time assessment of ventricular volumes and function enables data acquisition during free-breathing. However, in children the requirement for high spatiotemporal resolution requires accelerated imaging techniques. In this study, we implemented a novel real-time bSSFP spiral sequence reconstructed using Compressed Sensing (CS). The purpose of this study was to validate this real-time sequence against the breath-hold (BH) reference standard for assessment of ventricular volumes in children with heart disease.

**Methods**

Data was acquired in 60 children. Qualitative image scoring and evaluation of ventricular volumes was performed by three clinical cardiac MR specialists. 30 cases were reassessed for intra-observer variability, and the other 30 cases reassessed for inter-observer variability.

**Results**

Spiral real-time images were of good quality, however qualitative scores reflected more residual artefact than standard BH images ($p<0.0001$) and slightly lower edge definition ($p=0.0004$).

Quantification of both Left Ventricular (LV) and Right Ventricular (RV) metrics showed excellent correlation between the techniques with narrow limits of agreement. However, we observed small but statistically significant overestimation of LV end-diastolic volume (EDV, of 1.7%) as well as underestimation of LV end-systolic volume (ESV, of 1.0%) as well as in calculated metrics using these values using the real-time imaging technique. We also observed a small overestimation of RV stroke volume (SV, of 1.8%) and ejection fraction (EF, of 1.4%) using the real-time imaging technique. No difference in inter-observer or intra-observer variability were observed between the BH and real-time sequences.

**Conclusions**

Real-time bSSFP imaging using spiral trajectories combined with a compressed sensing reconstruction is feasible. Despite slightly lower image quality compared to standard BH technique, we saw a good agreement for quantification of both LV and RV metrics in children with heart disease. The main benefit over standard BH imaging is that it can be acquired during free breathing. However, another important secondary benefit is that a whole ventricular stack can be acquired in approximately 20 seconds, as opposed to almost 6 minutes for standard BH imaging. Thus, this technique holds the potential to significantly shorten MR scan times in children.

**BACKGROUND**

Evaluation of ventricular volumes and function is vital in the investigation of children with heart disease. The reference standard method of evaluating ventricular volumes is cardiovascular magnetic resonance (MR) – specifically, multi-slice cardiac gated balanced steady state free precession (bSSFP) cine imaging [1]. This technique is now routinely used in children and has an important role to play in disease management [2]. Unfortunately, it does require multiple breath-holds and this can be difficult in young children with heart disease.

An alternative approach is real-time MR in which multiple *k*-space frames are acquired within a single heartbeat. The benefit of real-time MR is that it can be acquired during free breathing and without cardiac gating. However, this comes at the cost of lower spatial and temporal resolution, which can affect accuracy [3]. One solution is to leverage accelerated imaging techniques (data undersampling or more efficient *k*-space filling) to improve resolution.

Compressed sensing (CS) is currently one the most powerful methods of reconstructing heavily undersampled data [4]. It is often combined with radial *k*-space filling to achieve incoherent sampling [4, 5]. Previous studies in adults have demonstrated good agreement between radial real-time techniques and the breath-hold reference standard for assessment of ventricular volumes [6-9]. However, imaging children requires much higher spatial and temporal resolution, hence greater acceleration is needed to translate these techniques into the paediatric population. One possibility is to combine CS with more efficient spiral *k*-space filling. Spiral trajectories have previously been used to accelerate real-time phase contrast MR with some success [10]. They can also be combined with golden angle spacing [11, 12] to enforce incoherent aliasing necessary for CS reconstruction.

In this study, we implemented a novel real-time bSSFP spiral sequence reconstructed using CS. The purpose of this study was to validate this real-time sequence against the breath-hold reference standard for assessment of ventricular volumes in children with heart disease.

**METHODS**

**Study Population**

Between December 2017 and February 2018, 60 consecutive children referred for cardiac MR were enrolled into this study. The sample size was chosen to detect a mean difference in right ventricular end diastolic volume of 2 mL, assuming a standard deviation of the difference of 12 mL [3] and a statistical power of 80% with a P value of 0.05. This analysis gave an estimated sample size of 59 (which was increased to 60 for redundancy). The exclusion criteria were: a) general contraindications to MR, such as pregnancy or MR-incompatible implants, b) requirement for general anesthetic, c) irregular heart rate, d) difficulty performing breath-holds, and e) single ventricular anatomy. The local committee of the UK National Research Ethics Service approved the study (06/Q0508/124), and written consent was obtained from all subjects/guardians.

**Imaging Protocol**

All imaging was performed on a 1.5 Tesla MR scanner (Avanto, Siemens Medical Solutions, Erlangen, Germany) with maximum gradient amplitude of 40 mT/m and a maximum slew rate of 180 T/m/s. Two spine coils and one body-matrix coil (giving a total of 12 coil elements) were used for signal detection and a vector electrocardiographic system was used for cardiac gating.

*Standard volumetric assessment* — Standard cardiac-gated ventricular volume assessment was performed using a multi-slice retrospectively cardiac gated bSSFP Cartesian sequence. The temporal resolution was 29.5 ms and the in-plane spatial resolution was 1.5x1.5 mm (full sequence parameters are shown in table 1). The imaging plane was in the ventricular short axis (SAX), which was planned using right and left ventricular long axes and 4-chamber images. Sufficient contiguous slices were acquired in the short axis to ensure coverage of the whole ventricle (~13±2 slices, range: 10 to 17). Each slice was acquired in a separate breath-hold, each lasting ~5.5±1.1 seconds (range: 3.6 to 8.8 seconds).

|  | **Standard BH technique** | **Spiral real-time technique** |
|---|---|---|
| Field of view (mm) | 350 | 350 |
| Rectangular field-of-view (%) | 75 | 100 |
| Matrix size | 240 x 180 | 208 x 208 |
| Number of slices | ~13 (range: 10-17) | ~13 (range: 10-17) |
| Image resolution (mm) | 1.46 x 1.46 x 8 | 1.68 x 1.68 x 8 |
| TE/TR (ms) | 1.16 / 2.32 | 0.67 / 3.34 |
| Flip Angle (degree) | 67 | 67 |
| Pixel Bandwidth (Hz/pixel) | 1225 | 1502 |
| Cardiac gating | Retrospective | - |
| Temporal resolution (ms) | 29.48 | 30.06 |
| Reconstructed cardiac phases | 40 | ~24 (range: 20-37) |
| Trajectory | Cartesian | Spiral |
| Fully sampled spiral interleaves | - | 72 |
| GRAPPA | x2 | - |
| Compressed Sensing | - | x8 |
| Breath-hold time per slice (s) | 3.6-8.8 | - |
| Total acquisition time for SAX (s) | ~350 (range: 239-573) | ~20 (range: 11 to 36) |

*Table 1: Sequence parameters*

*Real-time volumetric assessment* — Real-time bSSFP imaging was performed using a modified uniform density spiral sequence. A zeroth moment rewinder gradient was added at the end of the spiral readout to maintain bSSFP coherence. However, in this study we did not use bipolar first moment rewinder gradients, due to the associated increase in TR [13]. The spiral trajectories were designed using the method described by Hargreaves [14], assuming that 72 regularly spaced spiral interleaves were needed to completely fill *k*-space (each one of duration 1.86 ms). Consecutive interleaves were rotated by the tiny golden angle (tGA, ~47.26°) rather than the golden angle to reduce eddy currents associated with large *k*-space jumps [15]. An acceleration factor of 8 was used, resulting in 9 spiral interleaves per frame. Real-time cine acquisition of each slice was acquired over two R-R intervals, with the first R-R interval used to reach the steady state and the second for data acquisition. The imaging plane and the number of slices was the same as the standard BH sequence. All real-time image acquisition was performed during free breathing. The temporal resolution was 30.1 ms and the in-plane spatial resolution was 1.7x1.7 mm (full sequence parameters are shown in table 1).

The CS reconstruction was performed on an external GPU equipped computer (Tesla K40c, NVIDIA) with on-line communication to the native reconstructor [16, 17]. Sparsity in the CS reconstruction was enforced using a temporal finite difference operator, which is similar to the previously described iterative Golden-angle Radial Sparse Parallel (GRASP) reconstruction [18]. All regularization parameters were selected empirically on the first 10 patients, to optimize artefact removal without temporal blurring. Additionally, coil sensitivity information was calculated from the time-averaged data (from each slice) [19].

**Image Analysis**

Qualitative image scoring and evaluation of ventricular volumes was performed by three clinical cardiac MR specialists (OT – 12 years experience, MH – 11 years experience, and KM – 5 years experience). Each clinician was primary observer for 20 unique cases, of which 10 were re-evaluated to assess intra-observer variability. In addition, each observer assessed 10 cases from a different primary observer (5 cases from each of the other observers), to evaluate inter-observer variability. Thus, each observer scored and processed 40 cases. Overall 30 cases were used to evaluate intra-observer variability and the other 30 cases used to evaluate inter-observer variability.

*Image Quality* – The mid-ventricular short-axis cine loops from each technique were scored on a 5-point Likert scale in four categories: sharpness of endocardial border (1 = non-diagnostic, 2 = poor, 3 = adequate, 4 = good, 5 = excellent), temporal fidelity (or blurring) of wall motion (1 = non-diagnostic, 2 = poor, 3 = adequate, 4 = good, 5 = excellent) and residual artefacts (1 = non-diagnostic, 2 = severe, 3 = moderate, 4 = mild, 5 = minimal). Each observer scored the patient data that they had previously segmented (N = 40 cases each), at a separate session. All loops were presented in a random manner using a custom-built python application, with the observers blinded to diagnosis, patient number and type of sequence.

*Ventricular Function* – Quantification of left ventricular (LV) and right ventricular (RV) volumes was performed in a similar manner for each technique using the OsiriX open source DICOM viewing platform (OsiriX v.9.0, OsiriX foundation, Switzerland) [20]. Firstly, the end-diastolic and end-systolic phases were identified for each ventricle through visual inspection of the mid-ventricular cine. The endocardial borders of all slices at end systole and diastole were then traced manually (including papillary muscles and trabeculation in the myocardial mass). This allowed calculation of end-diastolic volume (EDV) and end-systolic volume (ESV). Stroke volume (SV) was obtained by subtracting ESV from EDV and ejection fraction (EF) = SV/EDV × 100. In addition, LV epicardial borders were traced in end-diastole and combined with endocardial borders to obtain LV mass. Observers were presented with each anonymized volume in a random order, blinded to diagnosis, patient number and type of sequence.

**Quantitative image assessment**

Calculation of signal-to noise ratio (SNR) and contrast-to-noise ratio (CNR) in images reconstructed using CS is nontrivial due to the uneven distribution of noise. Therefore, we calculated blood pool-to-myocardial signal intensity ratio [3] as a quantitative measure of image contrast, which is important when segmenting data. In all patients, the average blood pool and myocardial signal intensity values were measured in the end diastolic volume (from the LV endocardial and LV epicardial borders drawn for volumetric calculation, across the entire volume). The image contrast equalled blood pool signal intensity divided by myocardial signal intensity.

Edge sharpness was calculated by measuring the maximum gradient of the normalized pixel intensities across the border of the septum, as previously described [21]. To reduce noise, which results in artificially high gradients (representing sharp edges), the pixel intensities were fit to a tenth order polynomial, before differentiation. Edge sharpness was calculated in six positions across the septum, for all cardiac phases, and the average value was used for comparison.

**Statistics**

Statistical analyses were performed by using STATA software (STATA SE, v.14.2). All results are expressed as the mean ± standard deviation. Mean ventricular volumes, function and mass measured using the standard BH and spiral real-time techniques were compared using a paired t-test. For assessment of agreement of ventricular volumes and function, the standard BH data was used as the reference standard for Bland-Altman analysis. Inter and intra-observer variability was assessed using the coefficient of variation (CoV) calculated using the root mean method. The CoV's for the standard BH and spiral real-time techniques were compared using the paired t-test. In addition, inter-obeserver and intra-observer variability were also assessed using one-way intraclass correlations (ICC). Qualitative image scores were also tested using the paired t-test. This was done as previous work has shown that the paired t-test has a lower Type II error rate compared to non-parametric tests for Likert scale data [22]. It is therefore more likely to detect differences between the two techniques. A p-value of less than 0.05 indicated a significant difference.

**RESULTS**

**Demographics and feasibility**

The mean age of population was 13.6±2.7 years (median: 13.5 years, range: 7.0-18.3 years) and 33 were female. The mean heart rate was 82.3±15.9 bpm (median: 82 bpm, range: 50-114 bpm). The full demographic information and patient diagnoses are shown in table 2. Real-time and standard BH imaging was successfully performed in all subjects. The total time acquisition time for the full BH stack was 348±79 seconds (median: 335 s, range: 239-573 s) compared to 19.8±5.8 seconds for the real-time spiral stack (median: 19 s, range: 11-36 s). The online CS reconstruction for each real-time slice was ~1.5s.

|  | **Mean ± standard deviation** |
| --- | --- |
| Male/Female | 27 / 33 |
| Age (years) | 13.6 ± 2.7 (range: 7.0 to 18.3) |
| Height (m) | 1.6 ± 0.2 (range: 1.2 to 1.9) |
| Weight (kg) | 47.5 ± 15.8 (range: 23.1 to 82.0) |
| BSA | 1.4 ± 0.3 (range: 0.9 to 2.0) |
| Heart rate (bpm) | 82 ± 16 (range: 50 to 114) |
| Pulmonary Hypertension | 11 |
| Cardiomyopathy | 10 |
| Family history Sudden death/cardiomyopathy | 9 |
| Pulmonary artery stenosis | 3 |
| Coarctation of the aorta | 4 |
| Tetralogy of Fallot (corrected) | 4 |
| Transposition of the great arteries (corrected) | 3 |
| Atrial septal defect | 3 |
| Ventricular Septal Defect | 2 |
| Aortopathy | 5 |
| Left ventricular outflow tract obstruction | 3 |
| Atrio-ventricular valve dysfunction | 3 |

*Table 2: Full demographic information and patient diagnoses*

**Image quality**

Representative images are shown in figures 1 and 2, and the corresponding movies can be seen in supplemental materials 1-4. These demonstrate that the spiral real-time images are of good quality. Nevertheless, qualitative image scores reflect that the spiral real-time images have more residual artefact than standard BH images (artefact score: 3.8±0.1 vs 4.8±0.1, p<0.0001) and have slightly lower edge definition (edge score: 4.4±0.7 vs 4.8±0.5, p=0.0004). There was also a trend towards slightly poorer motion fidelity (motion score: 4.6±0.1 vs 4.8±0.1, p=0.06). There were no significant differences between qualitative image scores for the different observers (p>0.1).

Quantitative edge sharpness was also significantly lower (p<0.0001) for the spiral real-time images compared to the standard BH images (0.68±0.19 mm$^{-1}$ vs 0.53±0.16 mm$^{-1}$, respectively). Additionally, quantitative estimates of image contrast (blood pool-to-myocardial signal intensity ratio) were significantly lower (p<0.0001) for the spiral real-time images compared to the standard BH images (2.7±0.4 vs. 3.2±0.4, respectively).

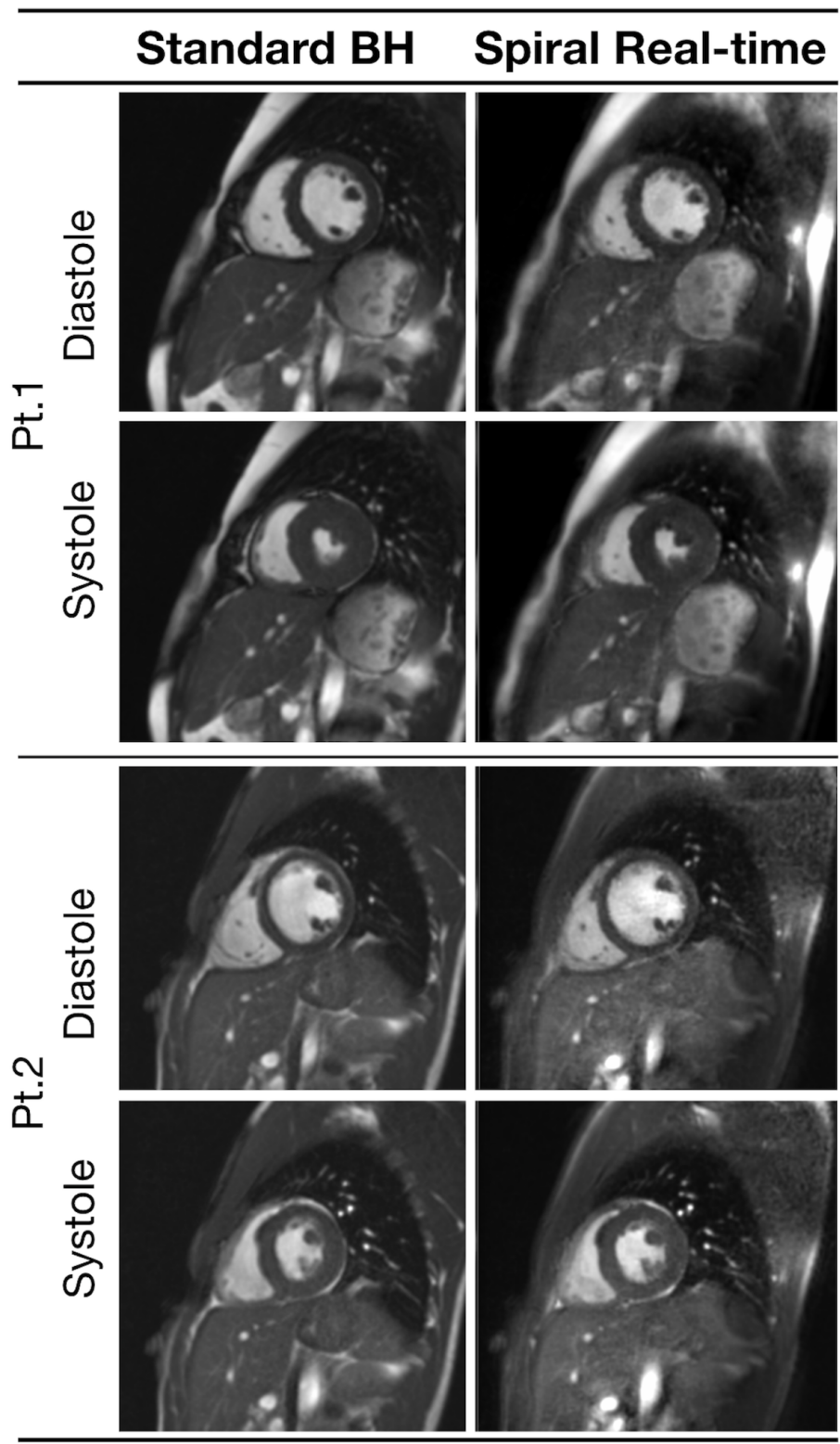

*Figure 1: Example image quality from two patients with left heart disease.*

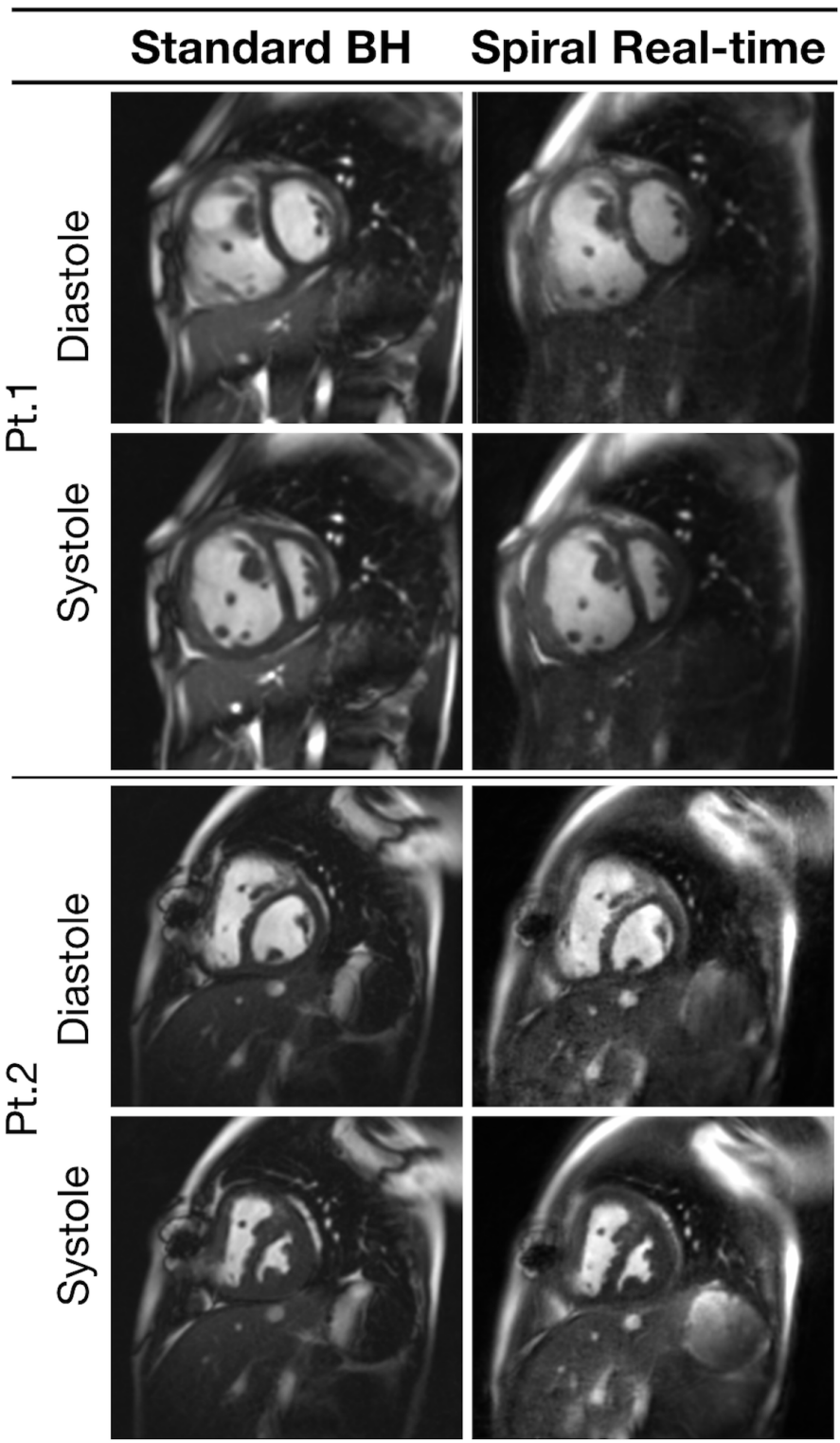

*Figure 2: Example image quality from two patients with right heart disease.*

**Ventricular volume quantification**

Ventricular metrics measured using spiral real-time and standard BH imaging are shown in table 3. Bland-Altman plots for LV metrics are shown in figures 3 and 4. There was a small (1.7%) but statistically significant overestimation of LV EDV, and a small (1.0%) but statistically significant underestimation of LV ESV using the real-time imaging technique. This resulted in small overestimations of LV SV (2.6%) and LVEF (1.5%). There was also a small but significant underestimation of LV mass (3.1%). Nevertheless, the limits of agreement for all LV metrics were narrow with excellent correlation (r≥0.90) between real-time and BH data.

Bland-Altman plots for RV metrics are shown in figures 4 and 5. There were no significant differences in RV EDV or ESV, however there was a small overestimation of RV SV (1.8%) and RV EF (1.4%). However, the limits of agreement for all RV metrics were narrow with excellent correlation (r≥0.93) between real-time and BH data.

|  | Standard BH ‡ | Real-time ‡ | Bias† | Lower limit agreement† | Upper limit agreement† |
|---|---|---|---|---|---|
| LV EDV (mL) | 110 ± 38 | 112 ± 38 * | 1.7 (0.5 to 2.9) | -7.3 (-9.3 to -5.2) | 10.7 (8.6 to 112.7) |
| LV ESV (mL) | 40 ± 16 | 39 ± 16 * | -1.0 (-1.9 to -0.1) | -7.9 (-9.5 to -6.4) | 5.9 (4.3 to 7.5) |
| LV SV (mL) | 70 ± 25 | 73 ± 24 * | 2.6 (1.4 to 3.9) | -6.7 (-8.9 to -4.6) | 12.0 (9.9 to 14.1) |
| LV EF (%) | 64 ± 6 | 66 ± 7 * | 1.5 (0.7 to 2.3) | -4.3 (-5.7 to -3.0) | 7.3 (6.0 to 8.7) |
| LV mass | 88 ± 36 | 85 ± 34 * | -3.1 (-5.0 to -1.3) | -16.9 (-20.1 to -13.8) | 10.6 (7.5 to 13.8) |
| RV EDV (mL) | 121 ± 37 | 121 ± 38 | 0.5 (-0.9 to 1.8) | -9.6 (-12.0 to -7.3) | 10.6 (8.3 to 12.9) |
| RV ESV (mL) | 51 ± 23 | 49 ± 24 | -1.2 (-2.5 to 0.0) | -11.0 (-13.2 to -8.7) | 8.5 (6.3 to 10.7) |
| RV SV (mL) | 70 ± 20 | 72 ± 20 * | 1.8 (0.5 to 3.1) | -7.8 (-10.0 to -5.6) | 11.4 (9.2 to 13.6) |
| RV EF (%) | 59 ± 9 | 61 ± 10 * | 1.4 (0.4 to 2.4) | -6.0 (-7.6 to -4.3) | 8.7 (7.0 to 10.4) |

‡ *Displayed as mean ± standard deviation*
† *Displayed as mean (95% confidence intervals)*
\* *Indicates significant differences with Standard BH technique (p<0.05)*

*Table 3: Primary observer; ventricular metrics measured using real-time and standard BH imaging*

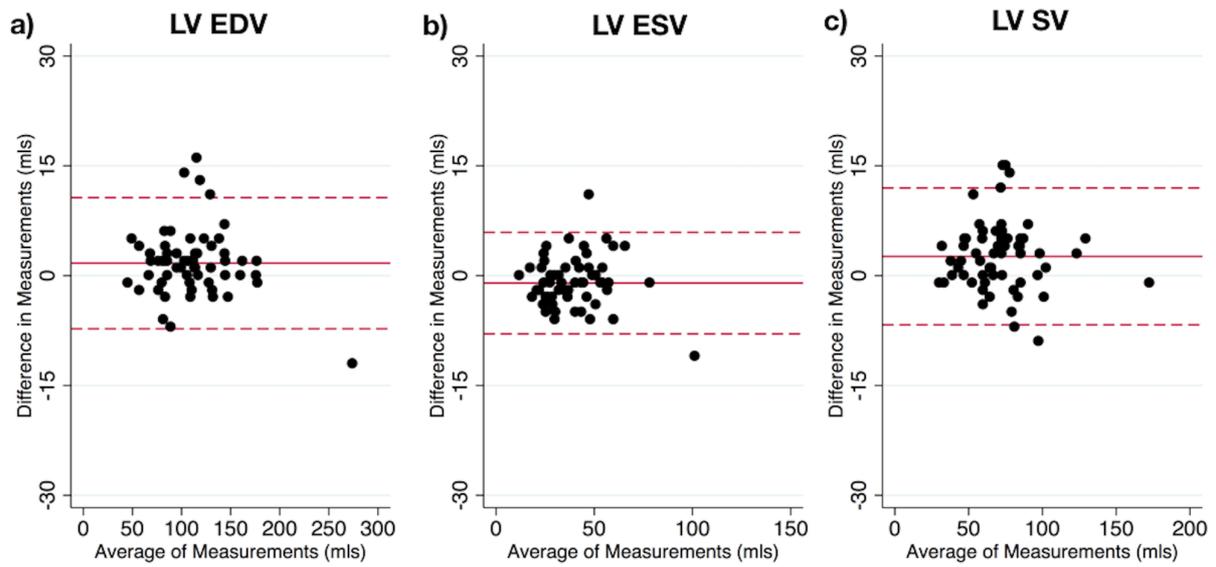

*Figure 3: Bland-Altman plots of standard BH vs real-time techniques for left ventricular a) EDV, b) ESV, c) SV.*

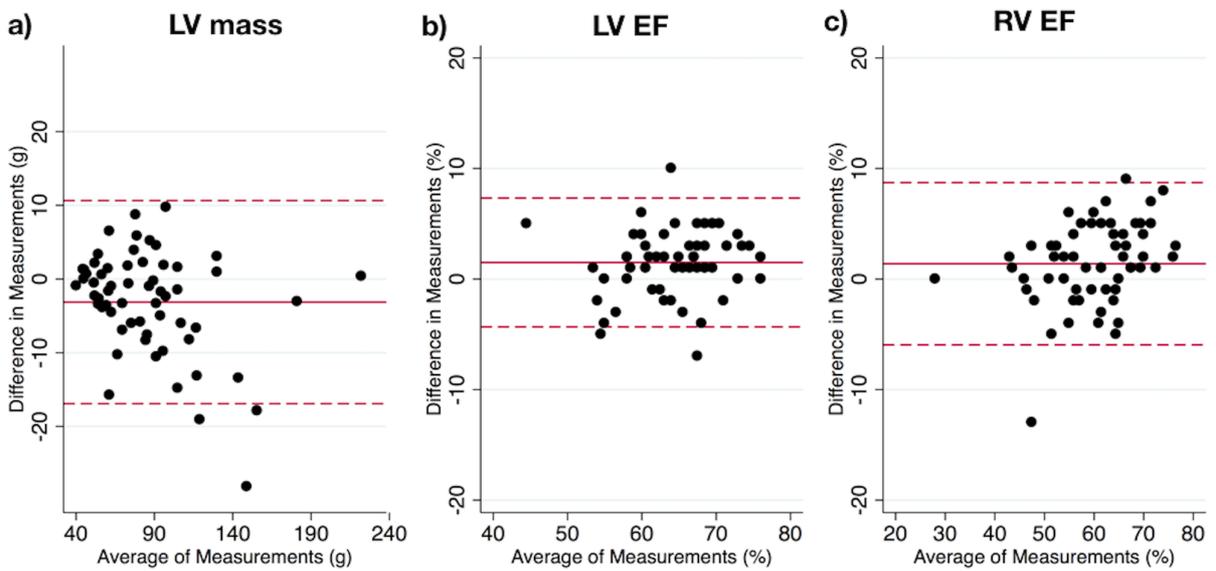

*Figure 4: Bland-Altman plots of standard BH vs real-time techniques for; a) left ventricular mass, b) left ventricular EF, c) right ventricular EF.*

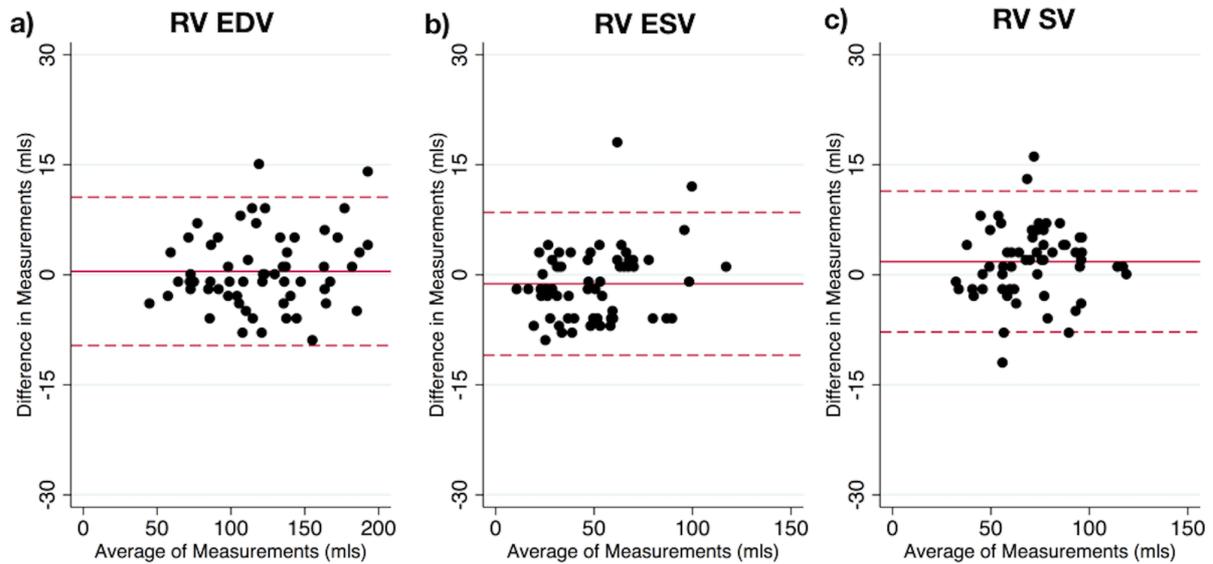

*Figure 5: Bland-Altman plots of standard BH vs real-time techniques for right ventricular; a) EDV, b) ESV, c) SV.*

The inter-observer CoV's and ICC's are shown in table 4 and intra-observer CoV's and ICC's are shown in table 5. There were no significant differences in inter-observer and intra-observer variability between spiral real-time and standard BH derived RV and LV metrics.

|  | **Standard BH CoV ‡** | **Real-time CoV ‡** | **CoV p-value** | **Standard BH ICC †** | **Real-time ICC †** |
|---|---|---|---|---|---|
| LV EDV (%) | 4.0 ± 4.5 | 3.9 ± 4.6 | 0.836 | 0.984 (0.968 to 0.992) | 0.983 (0.964 to 0.993) |
| LV ESV (%) | 6.8 ± 8.4 | 8.3 ± 12.8 | 0.487 | 0.931 (0.862 to 0.967) | 0.973 (0.946 to 0.967) |
| LV SV (%) | 6.9 ± 7.5 | 6.6 ± 7.3 | 0.762 | 0.952 (0.902 to 0.977) | 0.958 (0.914 to 0.977) |
| LV EF (%) | 4.1 ± 5.0 | 5.5 ± 9.6 | 0.435 | 0.772 (0.578 to 0.884) | 0.891 (0.785 to 0.886) |
| RV EDV (%) | 6.3 ± 6.2 | 6.7 ± 6.9 | 0.503 | 0.956 (0.911 to 0.979) | 0.964 (0.926 to 0.983) |
| RV ESV (%) | 12.3 ± 14.1 | 12.0 ± 13.8 | 0.813 | 0.954 (0.906 to 0.978) | 0.968 (0.935 to 0.985) |
| RV SV (%) | 10.3 ± 11.2 | 10.8 ± 11.9 | 0.560 | 0.835 (0.684 to 0.917) | 0.865 (0.738 to 0.933) |
| RV EF (%) | 6.9 ± 7.8 | 7.4 ± 7.9 | 0.299 | 0.785 (0.599 to 0.891) | 0.836 (0.687 to 0.918) |

‡ *Displayed as mean ± standard deviation*

† *Displayed as mean (95% confidence intervals)*

*Table 4: Inter-observer variability; ventricular metrics measured using real-time and standard BH imaging*

|  | **Standard BH CoV** ‡ | **Real-time CoV** ‡ | **CoV p-value** | **Standard BH ICC** † | **Real-time ICC** † |
|---|---|---|---|---|---|
| LV EDV (%) | 1.7 ± 2.6 | 1.5 ± 2.1 | 0.643 | 0.997 (0.994 to 0.999) | 0.994 (0.987 to 0.997) |
| LV ESV (%) | 4.2 ± 4.9 | 4.6 ± 5.5 | 0.620 | 0.984 (0.967 to 0.992) | 0.985 (0.968 to 0.993) |
| LV SV (%) | 2.3 ± 2.7 | 2.5 ± 3.7 | 0.698 | 0.994 (0.988 to 0.997) | 0.993 (0.985 to 0.997) |
| LV EF (%) | 1.8 ± 2.5 | 2.4 ± 3.2 | 0.290 | 0.919 (0.839 to 0.961) | 0.955 (0.909 to 0.978) |
| RV EDV (%) | 2.0 ± 2.9 | 2.1 ± 2.7 | 0.717 | 0.994 (0.987 to 0.997) | 0.994 (0.987 to 0.997) |
| RV ESV (%) | 4.1 ± 5.8 | 4.8 ± 6.9 | 0.402 | 0.977 (0.952 to 0.989) | 0.989 (0.977 to 0.995) |
| RV SV (%) | 3.5 ± 5.5 | 4.1 ± 6.0 | 0.305 | 0.985 (0.968 to 0.993) | 0.989 (0.977 to 0.995) |
| RV EF (%) | 2.6 ± 4.7 | 3.3 ± 4.6 | 0.161 | 0.946 (0.890 to 0.974) | 0.975 (0.949 to 0.988) |

‡ *Displayed as mean ± standard deviation*

† *Displayed as mean (95% confidence intervals)*

*Table 5: Intra-observer variability; ventricular metrics measured using real-time and standard BH imaging*

**DISCUSSION**

The main findings of this study were that: i) real-time bSSFP imaging using tiny golden-angle spiral trajectories combined with a compressed sensing reconstruction was feasible, ii) the image quality of the real-time technique was slightly lower than the standard BH technique, iii) there was good agreement for quantification of both LV and RV metrics between the spiral real-time and BH reference standard techniques.

**Spiral bSSFP sequence with CS reconstruction**

Real-time CMR is particularly useful when imaging children, as it can be performed quickly and without breath-holds. In this study, we implemented a novel real-time sequence that leveraged both efficient spiral *k*-space filling and CS reconstruction. Spiral *k*-space filling has previously been used to successfully accelerate spoiled gradient echo sequences (i.e. phase contrast MR [23] and MR angiography [24]). However, there are some problems associated with combining spiral trajectories with bSSFP readouts [13]. These mainly relate to off-resonance effects and longer repetitions times. Previous studies have shown that these effects can be partly mitigated using both zeroth and first order moment rewinders to balance the imaging gradients [13, 25]. The problem with this approach is that it significantly reduces temporal scan efficiency. Therefore, we chose to use a shorter conventional zeroth order moment rewinder, combined with a relatively short readout (1.86 ms) to reduce off-resonance effects. However, this increases the number of interleaves required to fully sample *k*-space. Consequently, high acceleration factors were required to ensure adequate temporal resolution and we utilized CS to reconstruct artefact-free images. Our CS implementation shares many characteristics with the previously validated GRASP reconstruction for radial trajectories [18]. Specifically, we utilized tiny golden angle spacing to produce temporal incoherency and penalized temporal finite differences in the reconstruction. Combining spiral bSSFP with CS resulted in a real-time sequence with only slightly lower spatial resolution, and similar temporal resolution, to standard BH imaging.

**Clinical utility**

Clinical utility depends on the ability of the reader to process the images to accurately measure ventricular volumes. In this study, images were processed by three independent cardiac MR specialists (not involved in the development of the sequence) This is a more 'real-world' test of the sequence, in contrast to many studies in which a single observer (often involved in the development of the technology) is used.

We demonstrated that spiral real-time images were of diagnostic quality. However, they did have slightly lower edge definition and greater amount of residual artefact. This is not surprising as the real-time images had lower spatial resolution and were acquired with significant acceleration. Interestingly, although CS with temporal finite difference sparsity can result temporal blurring there was only a trend towards reduced motion fidelity.

Irrespective of differences in image quality, from a clinical point it is the ability to accurately measure ventricular volumes that is paramount. In this study, we observed small biases in RV ESV and LV EDV (as well as in calculated metrics using these values). Nevertheless, these were all below 5% and would be expected to have minimal effect on clinical decision making. More importantly, the limits of agreement were narrow, with excellent correlation between the techniques. Interestingly, the limits of agreement were smaller than previously described real-time methods (including those that utilize CS) [6, 26]. This was probably due to the higher spatial and temporal resolution provided by using spiral trajectories. In this study, we also examined inter-observer and intra-observer variability (half the populations tested for each). For the standard BH sequence the coefficients of variation and ICC's were similar to previously published data [27]. This was also true for the spiral real-time sequence, with no difference in inter-observer and intra-observer variability between the BH and real-time sequences. This is an important finding, as demonstrating reliability is vital for clinical translation. We believe that our findings show that spiral real-time imaging can successfully be used for assessment of paediatric heart disease. The main benefit over standard BH imaging is that it can be acquired during free breathing. Another important secondary benefit is that a whole ventricular stack can be acquired in approximately 20 seconds, as opposed to almost 6 minutes for standard BH imaging. Thus, this technique holds the potential to significantly shorten cardiac MR scan times in children.

**Limitations**

The main limitation of this study was the minimum age in the population was 7 years old, which limits the translatability of our findings to younger children. The reason for this age limit was that younger children would have found it difficult to perform the standard BH sequence necessary for validation. Nevertheless, future studies concentrating on optimization and validation of this technique for infants would be useful. Another limitation was that we did not assess RV mass in this study. Several studies have suggested that RV mass may be an important predictive marker in certain diseases [28]. However, RV mass is difficult to accurately quantify in patients without RV hypertrophy, even when using standard BH imaging Therefore, we did not include it this study, although it could be included in future studies concentrating on patient with RV disease. Finally, one disadvantage of compressed sensing is long reconstruction times. In this study, we used an online GPU based reconstruction that fed images back to the scanner in a clinically acceptable time frame (<30 seconds for the entire SAX). Nevertheless, more work is required so that reconstruction times can be reduced to that of standard Cartesian imaging.

## CONCLUSION

In conclusion, we implemented a novel spiral real-time bSSFP sequence with CS reconstruction. We showed good agreement between ventricular volumes measured using spiral real-time and standard BH techniques in children with heart disease. Thus, we believe that this technique could be used successfully in the paediatric population.